\documentclass[prd,a4paper,showpacs,byrevtex,preprint]{revtex4}
\usepackage{graphicx}
\usepackage{dcolumn}
\usepackage{amsmath}
\usepackage{array}
\usepackage{bm}
\usepackage{amssymb}
\usepackage{amsfonts}
\usepackage{youngtab}
\begin{document}
%\preprint{Preprint Universit\'{e} de Mons-Hainaut}

\title{Light baryon masses in different large-$N_c$ limits}

\author{Fabien \surname{Buisseret}}
\thanks{F.R.S.-FNRS Postdoctoral Researcher}
\email[E-mail: ]{fabien.buisseret@umons.ac.be}
\author{Claude \surname{Semay}}
\thanks{F.R.S.-FNRS Senior Research Associate}
\email[E-mail: ]{claude.semay@umons.ac.be}
\affiliation{Service de Physique Nucl\'{e}aire et Subnucl\'{e}aire,
Universit\'{e} de Mons--UMONS,
Acad\'{e}mie universitaire Wallonie-Bruxelles,
Place du Parc 20, B-7000 Mons, Belgium}

\date{\today}

\begin{abstract}
We investigate the behavior of light baryon masses in three inequivalent large-$N_c$ limits: 't~Hooft, QCD$_{{\rm AS}}$ and Corrigan-Ramond. Our framework is a constituent quark model with relativistic-type kinetic energy, stringlike confinement and one-gluon-exchange term, thus leading to well-defined results even for massless quarks. We analytically prove that the light baryon masses scale as $N_c$, $N_c^2$ and $1$ in the 't~Hooft, QCD$_{{\rm AS}}$ and Corrigan-Ramond limits respectively. Those results confirm previous ones obtained by using either diagrammatic methods or constituent approaches, mostly valid for heavy quarks. 
\end{abstract}


% 11.15.Pg    Large N_c
% 12.39.Ki    Relativistic quark model
% 12.39.Pn    Potential models
% 14.20.-c    Baryons

\maketitle

\section{Introduction}

The study of large-$N_c$ QCD finds its origin in the lack of obvious expansion parameter for QCD: Setting the number of colors equal to a large value $N_c$ instead of 3 allows indeed to make perturbative expansions in $1/N_c$~\cite{hoof}. The relevance of such a framework demands that a world in which the number of colors is very large should not be too different from our QCD world. The many successes of large-$N_c$ methods in understanding experimental as well as theoretical features of QCD provide an excellent \textit{a posteriori} justification of that statement~\cite{manohar}. Moreover, lattice and AdS/CFT-based calculations of various observables using different gauge groups strongly favor the idea that SU(3) is actually close to SU($\infty$) (see \textit{e.g.}~\cite{teper}). 

The standard way of generalizing QCD to a large number of colors is also the first one proposed in the literature, by 't~Hooft~\cite{hoof}. It consists in taking the quarks in the fundamental representation of SU($N_c$), and then letting the number of colors becoming arbitrarily large while the so-called 't~Hooft coupling $g^2 N_c$ remains constant for any $N_c$ ($g$ is the strong coupling constant). Note that in the 't~Hooft limit, the number of quark flavors remains finite; later Veneziano would propose to set $N_f=O(N_c)$~\cite{vene} so that the theory remains planar, but the internal quark loops are no longer suppressed as in the 't~Hooft limit. However, it has been noticed soon after 't~Hooft proposal that the extrapolation of QCD to arbitrary $N_c$ is not unique: The only theoretical constraint is indeed that the resulting theory reduces to QCD at $N_c=3$~\cite{corr}. In this last reference for example, it has been suggested that some quark flavors could be in the $A_{N_c-2}$ representation, that is the $(N_c-2)$-indices antisymmetric representation, equal to the fundamental one at $N_c=3$. This is called the Corrigan-Ramond limit. Moreover, since the fundamental and $A_2$ representations are equivalent for SU(3), another large-$N_c$ limit can be proposed in which quarks are taken to be in the $A_2$ representation of SU($N_c$). Denoted QCD$_{{\rm AS}}$, that limit interestingly leads to a theory equivalent to ${\cal N}=1$ SUSY Yang-Mills, as shown in \cite{qcdas}.

In any large-$N_c$ limit, mesons and glueballs are always quark-antiquark and gluonic bound states respectively, but baryons need to be considered more carefully. As pointed out by Witten~\cite{witten}, baryons in 't~Hooft limit should be seen as bound states made of $N_c$ quarks in a totally antisymmetric color singlet. In that reference, using Feynman diagrams and a nonrelativistic quark model, it has been shown that the baryon masses scale as $N_c$ at the dominant order. It was also suggested that the same result should hold for baryons made of light quarks. Remark that corrections in $1/N_c$ can be added to improve the agreement of baryonic mass formulas with experiment (see \textit{e.g.}~\cite{jenkins}). In the Corrigan-Ramond limit however, a baryon can be made of three quarks as in QCD, so its mass should stay constant with $N_c$ at dominant order~\cite{corr}. Finally, baryons in the QCD$_{{\rm AS}}$ should rather consist in $N_c(N_c-1)/2$ quarks in a totally antisymmetric color singlet~\cite{bolo}, with a mass scaling as $N_c^2$ as shown in \cite{cohen} by using diagrammatic methods similar to the ones of \cite{witten}.    

In the present paper, we propose a constituent quark model that has the peculiarity of being relevant in the light baryon sector: Quarks can be massless thanks to a kinetic term of relativistic form, and long-range confining interactions are taken into account within a flux-tube (or string) picture. Our model is discussed in Sec.~\ref{sec:bham}, where we keep the formalism general so that any baryonic system can be studied, either in QCD or in the aforementioned large-$N_c$ limits. Then, using mathematical tools that are detailed in Appendix~\ref{appendix}, we find analytical upper and lower bounds for the light baryon mass spectra in the 't~Hooft, QCD$_{{\rm AS}}$ and Corrigan-Ramond limits in Secs.~\ref{thooft}, \ref{aslim} and \ref{crlim} respectively. As finally summarized in Sec.~\ref{conclu}, those bounds allow to find how the light baryon masses scale with $N_c$ and provide a confirmation of results obtained in previous works devoted to that topic. 

\section{Baryonic Hamiltonians}\label{sec:bham}

\subsection{General considerations}

A Hamiltonian describing bound states of baryonic type is needed as a starting point for our study. Since the gauge group considered is SU($N_c$), and since the quarks are allowed to be in an arbitrary color representation $R$ of that group, we mean by ``baryonic type" a bound state made of quarks only, whose color function is a totally antisymmetric singlet with respect to the exchange of two quarks. Let us denote $n_q$ the minimal number of quarks necessary to build such a color singlet. Then the dominant term in the quark kinetic energy can be written under the spinless Salpeter form 
\begin{equation}
T=\sum_{i=1}^{n_q}\sqrt{\vec p^{\, 2}_i+m^2_i},
\end{equation}
which has the advantage of being well-defined even for massless quarks. It is sometimes called semirelativistic in the literature since it is not a covariant formulation.  

Let us now discuss the interactions between the quarks within a generic baryonic system. At the perturbative level, \textit{i.e.} at short distances, one-gluon-exchange processes are dominant. The corresponding short-range potential can be computed from the QCD Feynman diagrams at tree level. In the quark-quark case, one is led to the well-known Fermi-Breit interaction, whose leading part is spin-independent and of Coulomb form. Typically, one expects the one-gluon-exchange term to be
\begin{equation}\label{voge}
V_{oge}=\frac{1}{2}(C_{qq}-2C_q)\, \alpha_s\sum_{i<j=1}^{n_q}\frac{1}{\left|\vec x_i-\vec x_j\right|},
\end{equation}
where $\alpha_s=g^2/4\pi$ and where $C_{qq}$ and $C_q$ are the quadratic Casimir operators of SU($N_c$) in the representations of the quark-quark pair and of the quark respectively. $\vec x_i$ obviously denotes the position of quark $i$. Notice that we assumed that each quark pair in the system is in the same color channel. The reason of such a choice will appear more clearly in the following sections. As previously mentioned, the 't~Hooft large-$N_c$ limit is such that $N_c\rightarrow\infty$ and $\alpha_s=\alpha_0/N_c$, with $\alpha_0$ independent of $N_c$. Since $\alpha_s\approx0.35$-0.40 in various potential models of real QCD, $\alpha_0\approx 1$. With such a value, all arguments within the square roots appearing in the following to compute baryon masses are well definite positive numbers. 

The long-range part of the interactions, responsible for the confinement, will be described within the framework of the flux-tube model. From the Casimir scaling hypothesis, one considers that each color source generates a straight flux tube whose energy density (or tension) is proportional to its quadratic color Casimir operator. Then, the flux tubes have to meet in one or several points such that the total energy contained in those flux tubes is minimal. In ``QCD" baryons (three quark systems), the junction point is identified with the Steiner (or Fermat or Toricelli) point of the triangle made by the quarks, leading to a Y-junction configuration as observed in lattice computations~\cite{ichie,taka,bissey}. Excepted for highly asymmetric quark configurations, that are not expected to be relevant for low-lying baryons, the junction point can be identified to the center of mass of the system within an excellent approximation~\cite{bsb}. Inspired by the usual baryonic case, we will assume a confining potential of the form
\begin{equation}\label{vc}
V_c=\frac{C_q}{C_{\yng(1)}}\sigma\sum_{i=1}^{n_q}\left|\vec x_i-\vec R\right|,
\end{equation}
where $\vec R$ is the center of mass of the system. According to the Casimir scaling, $(C_q/C_{\yng(1)})\sigma$ is the flux-tube energy density since $C_{\yng(1)}$ is the quadratic Casimir in the fundamental representation and $\sigma$ is the fundamental string tension, that is the one between a quark-antiquark pair. It is expected that $\sigma$ is constant with $N_c$ at leading order \cite{make}. In various potential models, $\sigma\approx 0.15$-0.20~GeV$^2$. Actually, there may be more than one junction point for multiquark states (tetraquarks, pentaquarks, \dots), and algorithms exist to compute them (see~\cite{Bicu3} as well as lattice computations~\cite{multiq}). So, a baryon made of $n_q$ quarks is probably a very complicated spatial structure with numerous strings connecting quarks and multiple junctions in order to minimize the potential energy. Nevertheless, besides the fact that potential~(\ref{vc}) is of one-body form, which is better in view of analytical calculations, it has several interesting physical features:  
\begin{itemize}
\item When $N_c=3$, it reduces to an excellent approximation of the genuine Y-junction.
\item Provided that a single junction point is present, its motion can be neglected at large-$N_c$ (see arguments in \cite{witten}) and its position can thus be identified with the center of mass.
\item At large-$N_c$, the string tension only depends on the $N$-ality of the representation (in our case, the number of fundamental representations needed to build a given representation by tensor product). This has been shown using several approaches such as holographic techniques or lattice calculations \cite{armo1,armo2}. If $k$ fundamental strings emerging from $k$ quarks in the fundamental representation connect to a junction, the energy density of the string emerging from this junction is expected to be $\sigma_k=k\,\sigma$ at the dominant order in $N_c$. The corresponding energetic cost is similar to the one for $k$ fundamental strings. So, the connection of the strings emerging from all quarks in the baryon into a single point is not energetically disadvantaged with respect to the apparition of multiple junctions inside the baryon at large-$N_c$.
\end{itemize}
For these reasons, we will keep potential~(\ref{vc}) in our calculations, since it seems to be relevant for both $N_c=3$ and $N_c\to \infty$. 

It is worth mentioning that the total Hamiltonian, defined by 
\begin{equation}\label{hb}
H_b=T+V_{oge}+V_c	,
\end{equation}
is very similar to the string model of light baryons at large $N_c$ proposed by Witten in his seminal paper~\cite{witten}. The peculiarity of the present work is that the model will be generalized to other large-$N_c$ limits than 't~Hooft's one and that we use techniques allowing to obtain approximate analytical mass spectra for $H_b$. A last comment that can be done about our model concerns the strong coupling constant. Already at one-loop, $\alpha_S$ is running with the energy scale $\rm q^2$ and we know that $\alpha_s(\rm q^2\rightarrow\infty)=0$. Moreover, we consider that $\alpha_s(\rm q^2\rightarrow0)$ tends to a finite, nonzero value that corresponds to $\alpha_s$ used in~(\ref{voge}). This is coherent with the aforementioned lattice QCD calculations, showing in general that the static potential between quarks can be accurately fitted by a potential separated into a dominant nonperturbative part (the flux tubes) and a residual perturbative part (the one-gluon-exchange potential with a ``frozen" value of $\alpha_s$). 

The spin effects will be completely ignored in this work. For instance in the real world, the ratio $(m_\Delta-m_N)/\frac{1}{2}(m_\Delta+m_N)$, which is attributed to spin (possibly isospin) dependent interactions, is 27\% (around $1/N_c$ with $N_c=3$). This spin contribution is not negligible but can be considered as a perturbation of order $1/N_c$ in the large-$N_c$ limit \cite{jenkins}. Previous works have considered the $1/N_c$ expansion for the mass operator in terms of various spin or isospin dependent operators inspired by simple constituent quark models \cite{pirj08,gale09}. Explicit calculations of the spin contribution have been calculated for $N_c=3$ in the framework of a nonrelativistic quark model \cite{gale09}. This problem for  a relativistic kinematics and arbitrary values of $N_c$ can be treated within our formalism and will be studied in a subsequent paper. Here, we only focus on the dominant effects in baryons.

\subsection{Upper and lower bounds}

Summarizing the discussion of the previous section, we are left with the baryonic Hamiltonian
\begin{equation}\label{hb1}
H_b=\sum_{i=1}^{n_q}\left[\sqrt{\vec p^{\, 2}_i}+a\left|\vec x_i-\vec R\right|\right]-\sum_{i<j=1}^{n_q}\frac{b}{\left|\vec x_i-\vec x_j\right|},
\end{equation}
\begin{equation}\label{abdef}
\hspace{-1.3cm}{\rm where}\quad	a=\frac{C_q}{C_{\yng(1)}}\sigma,\quad {\rm and} \quad b=\frac{1}{2}(2C_q-C_{qq})\frac{\alpha_0}{N_c}.
\end{equation}
Since we focus on light baryons, we have set $m_i=0$, which is a good approximation for the $u$ and $d$ quark masses.

Although finding the exact mass spectrum of $H_b$, denoted $M_b$, would require numerical computations, analytical techniques allow to find upper and lower bounds of that spectrum. For the sake of clarity, we leave the technical details for Appendix~\ref{appendix} while we give here the final results. From (\ref{mu1}) and (\ref{ml1}), we can deduce that the mass $M_b$ of a given eigenstate is such that    
\begin{equation}\label{bound}
 n_q\, \inf_{\left|\psi\right\rangle}\left\langle \psi\left|\sqrt{\vec p^{\, 2}}+\frac{n_q-1}{2}\left[\frac{a}{n_q}r-\frac{b}{r}\right]\right|\psi\right\rangle
 \leq M_b\leq\sqrt{4a(Q\, n_q-b\, P^{3/2})},
\end{equation}
where we recall that $P$ is the number of quark pairs and $Q$ is the band number of the considered state in a harmonic oscillator picture. The upper bound, obtained by the auxiliary field method (AFM), is state dependent, while the lower bound only concerns the ground state (hence, it is also a lower bound of the whole spectrum). Following (\ref{ml2}), it can be approximated by $\sqrt{2aP \left(2-b(n_q-1)\right)}$. The number of excitation quanta, $K=\sum_{i=1}^{n_q-1}(2n_i + \ell_i)$, could be seen as either of order 1 or of order $n_q$. We will consider in the following that $K=O(1)$ in order to get simpler mass formulas, but we point out that our final results would remain valid if $K=O(n_q)$. 

Whatever the values of $a$ and $b$, (\ref{bound}) and (\ref{QK}) implies that
\begin{equation}
M_b^2 \leq \alpha\, K + \beta,
\end{equation}
where $\alpha$ and $\beta$ do not depend on quantum numbers. So, for a fixed value of $n_q$, the upper bound predicts a Regge behavior for orbital and radial excitations. This is observed in the real QCD world with $N_c=3$. Moreover, large-$N_c$ methods help to understand why baryonic and mesonic Regge slopes are equal \cite{armo09}.

A mean field approximation has been used in various papers \cite{witten,bolo} to estimate the baryon masses in the large-$N_c$ limit. One can ask what are the connections of our model with this approach? The energy $\epsilon$ of a single quark in a central mean field with a funnel form $\alpha\, r-\beta/r$ is given by
\begin{equation}
\epsilon = \sqrt{4 \alpha (Q^*-\beta)} \quad \textrm{with} \quad Q=2 n^*+l^*+3/2,
\end{equation}
within the AFM approximation \cite{afms}. The mass $M^*$ of $n_q$ independent quarks is 
\begin{equation}
M^* = \sum_{i=1}^{n_q} \sqrt{4 \alpha (Q^*_i-\beta)}.
\end{equation}
For large values of $n_q$, we can assume a good equipartition of the energy $Q^*_i\approx Q/n_q$, where $Q$ is the number of quanta for the baryon state considered. The corresponding mass is then 
\begin{equation}
M^* \approx \sqrt{4 \alpha (Q\, n_q-\beta \, n_q^2)}.
\end{equation}
With the identification $\alpha=a$ and $\beta=2^{-3/2}b\, n_q$, the upper bound (\ref{bound}) is recovered. Nevertheless, a mean field approximation is not fully equivalent to our approach. For instance, in the mean field formulation, the wavefunction is the product of $n_q$ individual functions $\phi_j(\vec x_j -\vec R)$, which is very different from the wavefunction~(\ref{piphi}).

\section{'t~Hooft limit}\label{thooft}

In this case, the quarks are in the fundamental representation of SU($N_c$), and the number of colors becomes large while the 't~Hooft coupling, or equivalently $\alpha_s\, N_c$, remains constant for any $N_c$. The number of quark flavors is finite. The generalization of a baryon is then given by a state with $n_q=N_c$, the $N_c$ quarks forming a totally antisymmetric color singlet, implying that any quark pair is in the $A_2$ representation $\yng(1,1)$ \cite{witten}. Thus the Casimir operators read in this case
\begin{equation}
C_q=C_{\yng(1)}=\frac{N^2_c-1}{2N_c},\quad C_{qq}=C_{\yng(1,1)}=\frac{(N_c-2)(N_c+1)}{N_c}.	
\end{equation}
The interested reader can find in \textit{e.g.} \cite{cas} a way to compute the Casimir operators of SU($N_c$). This implies that the $a$ and $b$ factors~(\ref{abdef}) are now given by
\begin{equation}
a=\sigma,\quad b=\frac{N_c+1}{2N_c^2} \alpha_0.
\end{equation}
Due to the large value of $N_c$, one has $Q\rightarrow 3N_c/2$ ($K=O(1)$), $P\rightarrow N^2_c/2$, and (\ref{bound}) becomes
\begin{equation}\label{mb1}
 N_c\, \inf_{\left|\psi\right\rangle}\left\langle \psi\left|\sqrt{\vec p^{\, 2}}+\left(\frac{\sigma}{2}\, r-\frac{\alpha_0}{4\, r}\right)\right|\psi\right\rangle 
 \leq M_b\leq N_c\sqrt{\sigma\left( 6-\frac{\alpha_0}{\sqrt 2} \right)}.
\end{equation}
Following (\ref{ml2}), this last lower bound is close to the explicit formula $N_c\sqrt{\sigma  \left(2-\frac{\alpha_0}{2}\right)}$. Moreover, if $n_s$ strange quarks with nonzero mass are present, it is readily obtained from (\ref{dms}) that they bring a mass term approximately equal to
\begin{equation}\label{dms1}
\Delta M_s\approx n_s\frac{m^2_s}{6\sqrt{\sigma}}\sqrt{6-\frac{\alpha_0}{\sqrt 2}}.
\end{equation}
Using~(\ref{rad}), the mean square radius of the baryon, at the limit $N_c \to \infty$, is given by
\begin{equation}
\left\langle r^2\right\rangle \approx \frac{6-\sqrt{2}\alpha_0}{4 \sigma} .
\end{equation}
The size of the baryon tends toward a finite value in this limit, as suggested in \cite{witten,bolo}.

Since $\sigma$ and $\alpha_0$ do not depend on $N_c$, (\ref{mb1}) actually shows that any eigenvalue of Hamiltonian~(\ref{hb1}) scales as $N_c$ in the 't~Hooft limit. We argued in Sec.~\ref{sec:bham} that this last Hamiltonian provides a relevant description of baryonic systems, at least at the dominant order; consequently, we have shown that the light baryon masses scale as
\begin{equation}\label{mbscal1}
M_b\propto N_c
\end{equation}
in the 't~Hooft limit, just as it has been shown for heavy baryons in \cite{witten}. This is in agreement with the results in \cite{luty93} which show that this scaling must be verified whatever the number of light quark flavors. The strange quark contribution~(\ref{dms1}) is constant with $N_c$ and proportional to $m^2_s$, which is the SU$_F$(3)-breaking parameter in our model. That behavior is indeed coherent with former results obtained in large-$N_c$ baryon physics~\cite{jenkins,semay}. Moreover the mean square radius is also of order $O(1)$. 

We point out that, although $Q$ (or $K$) does not appear explicitly in the formulas anymore, these results are valid for any excited state of $H_b$ as well as for its ground state. It can finally be observed, from the formulas of Appendix~\ref{appendix}, that a pairwise confining potential would not have led to masses of order $N_c$ excepted if $\sigma$ was of order $1/N_c$, which seems not relevant as mentioned above. 

\section{QCD$_{{\rm AS}}$ limit}\label{aslim}

That limit is different from the 't~Hooft one since the quarks are now assumed to be in the $A_2$ representation $\yng(1,1)$ of SU($N_c$), demanding the baryonic color wave functions to be reconsidered. Indeed, the number of quarks needed to build a totally antisymmetric color singlet at any $N_c$ is now $n_q=N_c(N_c-1)/2$ as explicitly shown in \cite{bolo}. As for the 't~Hooft limit, $n_q$ is equal to the dimension of the quark color representation, meaning that all the quark color configurations appear only once in the singlet. If it is readily computed that $C_q=C_{\yng(1,1)}=(N_c-2)(N_c+1)/N_c$, it is nevertheless worth discussing a bit the value of $C_{qq}$. A quark pair inside the baryon can be in the following color channels:
\Yboxdim9pt
\begin{equation}\label{young}
\yng(1,1)\otimes\yng(1,1)=\yng(2,2)\oplus\, \yng(1,1,1,1)\oplus\, \yng(2,1,1)	.
\end{equation}
\Yboxdim5pt
However, the last channel is the only relevant one, for the following reasons. From a quark model point of view, one can check that $\yng(2,2)$ leads to repulsive interactions, that are thus not appropriate to describe a baryon. Moreover, the totally antisymmetric channel is forbidden from a diagrammatic point of view, as shown in \cite{cher}. The point is that a quark now carries two fundamental color indices. Consider a one-gluon-exchange process involving initially two quarks with indices [$ab$] and [$cd$]. The four indices must be different in order for the pair to be in the $A_4$ representation. The exchanged gluon would then change the pair into, say, [$ad$] and [$bc$]. But, in virtue of the Pauli principle, that pair is already present in the baryon, leading to the vanishing of that process. From that argument, the conclusion is that any quark pair must have only one common color index, typically [$ab$] and [$ac$]. This corresponds to the last diagram of (\ref{young}), and one finally concludes that $C_{qq}=C_{\yng(2,1,1)}=2(N_c^2-N_c-4)/N_c$.

The $a$ and $b$ factors are given, in the QCD$_{{\rm AS}}$ case, by
\begin{equation}
a=\frac{2(N_c-2)}{N_c-1}\sigma,\quad b=\frac{2}{N_c^2}\, \alpha_0,
\end{equation}
and, as $N_c\rightarrow\infty$, it can be checked that $Q\rightarrow 3N_c^2/4$ ($K=O(1)$) and $P\rightarrow N^4_c/8$. Equation~(\ref{bound}) becomes 
\begin{equation}\label{asbound}
 \frac{N_c^2}{2}\, \inf_{\left|\psi\right\rangle}\left\langle \psi\left|\sqrt{\vec p^{\, 2}}+\left(\sigma\, r-\frac{\alpha_0}{2\, r}\right)\right|\psi\right\rangle 
\leq M_b
\leq N_c^2\sqrt{\sigma\left(3-\frac{\alpha_0}{\sqrt 2} \right)}.
\end{equation}
Following (\ref{ml2}), that last lower bound is approximately equal to $N_c^2\sqrt{\sigma  \left(1-\frac{\alpha_0}{2}\right)}$. The strange quark contribution is moreover given by
\begin{equation}\label{ass}
\Delta M_s\approx n_s\frac{m^2_s}{6\sqrt{\sigma}}\sqrt{3-\frac{\alpha_0}{\sqrt 2}},
\end{equation}
while the baryon radius reads
\begin{equation}
\left\langle r^2\right\rangle \approx \frac{6-\sqrt{2}\alpha_0}{8 \sigma} .
\end{equation}

Using the same arguments as in the 't~Hooft limit, we can conclude from (\ref{asbound}) that we have shown that the light baryon masses scale as
\begin{equation}
M_b\propto N_c^2
\end{equation}
in the QCD$_{{\rm AS}}$ limit, as proposed in \cite{cher2} and proved in Witten's diagrammatic way in \cite{cohen}. Moreover, the SU$_F$(3)-breaking term brings a contribution independent of $N_c$ in this case also, as pointed out in \cite{cher2}. The behavior is similar for the baryon radius. Notice that, with quarks in the $A_2$ representation, a totally \emph{symmetric} color singlet could be built with $N_c$ quarks, that would lead to a baryon mass scaling in $N_c^{7/6}$ following \cite{bolo}. This can be understood within our framework also, because the spatial wave function should then be antisymmetric to preserve Pauli's principle. In that case, it can be computed within our framework that $Q\propto n_q^{4/3}$ at large $n_q$~\cite{afm}, leading to $M_b\propto N_c^{7/6}$ following (\ref{mu1}) and $\left\langle r^2\right\rangle\propto N_c^{1/3}$ following (\ref{rad}).
 
\section{Corrigan-Ramond limit}\label{crlim}

In that limit, baryons are three-quark states for any $N_c$: Two quarks are in the fundamental representation and one (say quark number 3) is in the $A_{N_c-2}$ one. The total interaction potential of our model reads, in the limit where $N_c\rightarrow\infty$ and where the 't~Hooft coupling is fixed, 
\begin{eqnarray}
V_{oge}+V_c&\rightarrow&\sigma \left(2|\vec x_3-\vec R|+|\vec x_1-\vec R|+|\vec x_2-\vec R|\right) %\nonumber\\&&	
-\frac{\alpha_0}{2}\left(\frac{1}{|\vec x_1-\vec x_3|}+\frac{1}{|\vec x_2-\vec x_3|}\right).
\end{eqnarray}

That potential, being constant with $N_c$, would lead to constant masses also since $n_q=3$. However, it is rather problematic since the baryon picture suggested is quite far from what is known at $N_c=3$. In particular, no one-gluon-exchange term is present between the quark 2 and 3: In the previous limits, the $1/N_c$ scaling of this term was compensated by the number of quarks scaling as $N_c$ at least, but here $n_q$ is finite. Moreover, the asymmetry of the confining potential would lead to a tower of excited states quite different from what is expected.  For those reasons, we prefer not to investigate more the Corrigan-Ramond limit. We mention for completeness that the phenomenology of the Corrigan-Ramond limit has already been studied in previous works~\cite{cher2,papa}, and found to be less realistic that the one coming from the 't~Hooft or QCD$_{{\ AS}}$ limits, in agreement with our discussion. Moreover another limit which is somewhat in between the 't~Hooft and Corrigan-Ramond ones has also been proposed \cite{rytt06}, but it is out of the scope of the present work since it requires a formalism in which quarks are Dirac spinors.

\section{Conclusions}\label{conclu}

In this paper, we have built a constituent model describing a baryonic system, that is a bound state made of quarks only. Our approach is a generalization of Witten's original proposal~\cite{witten}, \textit{i.e.} a one-gluon-exchange potential plus a stringlike confinement, but in which a kinetic term of relativistic form is added. Using analytical methods, we have been able to find analytical upper and (approximate) lower bounds for the baryon mass spectrum in the limit where quarks are massless. This is a peculiarity of the present work since analytical results are generally known only for heavy quarks. 

The most important point is that we have confirmed that $M_b = O(N_c)$ for light baryons in the 't~Hooft limit, and that $M_b = O(N_c^2)$ in the QCD$_{{\rm AS}}$ limit in agreement with the recent work~\cite{cohen}. Moreover, $M_b = O(1)$ in the Corrigan-Ramond limit, but this last case does not seem relevant for a constituent approach. We have checked that these results are obtained for the more general confining potential $\sum_{i=1}^{n_q} a\left|\vec x_i-\vec R\right|^p$ with $p>0$. But, whatever the value of $N_c$, baryon masses lie on Regge trajectories, for orbital and radial quantum numbers, only for $p=1$.

The contribution coming from strange quarks is of order $O(1)$ in the three considered cases, as suggested in~\cite{cher2}. It can also be checked that the typical baryon size is independent of $N_c$ for the three considered limits, recovering a result obtained within a Skyrme model of baryons~\cite{bolo}.

Those results can basically be summarized as follows: For light baryons, $M_b$ is proportional at the dominant order to the number of quarks needed to build a totally antisymmetric color singlet. This seems rather intuitive a posteriori but it was actually not trivial a priori for the following reasons. First, the quarks are massless, and the usual guess $M_b\propto n_q\, m_q$ is not applicable. Second, the explicit form of the potential does not allow to deduce such a scaling law. Third, a simple counterexample can be found as shown in Sec.~\ref{aslim}: If the color singlet is totally symmetric, the baryon masses do not scale as $n_q$ at large $n_q$, showing the nontrivial character of our result. 

\acknowledgments
The authors thank the F.R.S.-FNRS for financial support. We also thank Nicolas Matagne for helpfull discussions.

\begin{appendix}

\section{Useful formulas}\label{appendix}

In this appendix, we give analytical approximate mass spectra for a generic Hamiltonian of the form 
\begin{equation}\label{geneh}
H=\sum_{i=1}^{n_q}\left[\sqrt{\vec p^{\, 2}_i+m^2_i}+a_1\left|\vec x_i-\vec R\right|\right]
+\sum_{i<j=1}^{n_q}\left[a_2\left|\vec x_i-\vec x_j\right|-\frac{b}{\left|\vec x_i-\vec x_j\right|}\right],
\end{equation}
in view of an application to baryonic systems studied in the present paper. First of all, we focus on the ultrarelativistic limit $m_i=0$. Although standard quantum mechanical techniques have difficulties to handle that limit, the auxiliary field method (AFM) has been shown to be efficient in that case also. We refer the reader to \cite{afm} for a detailed discussion of that method, but for our purpose it is sufficient to mention that it leads to the following upper bound for the mass spectrum of Hamiltonian~(\ref{geneh})
\begin{equation}\label{mu1}
M^2_{u1}(Q)=4(a_1+a_2\, P^{1/2})(Q\, n_q-b\, P^{3/2}),
\end{equation}
with
\begin{equation}\label{QK}
Q=\frac{3}{2}(n_q-1)+K, \quad K=\sum_{i=1}^{n_q-1}(2n_i+\ell_i),\quad{\rm and}\quad P=\frac{n_q(n_q-1)}{2}.
\end{equation}
$Q$ is the band number (in a harmonic oscillator picture) for the eigenstates of $H$, and $P$ is the number of particle pairs. It is assumed that the value of $b$ is such that $M^2_{u1}$ is a positive number. In the case where $a_1=0$ or $a_2=0$, (\ref{mu1}) reduces to formulas previously found in \cite{afm}. Notice that, if $n_s$ ($< n_q$) quarks have a small nonzero mass $m_s$, the corresponding mass shift is given by~\cite{afm}
\begin{equation}\label{dms}
\Delta M_s(Q)\approx n_s\frac{m^2_s}{2\mu_0(Q)},	
\end{equation}
where $\mu_0(Q)$ is a massless quark's average kinetic energy, given by
\begin{equation}
\mu_0(Q)=Q \sqrt{\frac{a_1+a_2\, P^{1/2}}{Q\, n_q-b\, P^{3/2}}}=\frac{Q\,M_{u1}(Q)}{2(Q\, n_q-b\, P^{3/2})}.
\end{equation}

The AFM not only leads to approximate analytical mass spectra, but also to approximate analytical wave functions, allowing to compute various observables with a good accuracy~\cite{wave}. Combining results from \cite{afm,wave}, a $n_q$-body eigenstate is written
\begin{equation}\label{piphi}
\varphi=\prod^{n_q-1}_{j=1}\, \varphi_{n_j,\ell_j}(\lambda_j, \vec y_j),
\end{equation}
where $\varphi_{n_j,\ell_j}(\lambda_j, \vec y_j)$ is a three dimensional harmonic oscillator wave function, depending on the Jacobi coordinate $\vec y_j$ \cite{afm} and decreasing asymptotically like ${\rm e}^{-\lambda^2_j\, \vec y_j^{\, 2}/2}$. The quantum numbers are such that $\sum_{j=1}^{n_q-1}(2n_j+\ell_j)=K$ and the scale parameters $\lambda_j$ are given by
\begin{equation}\label{laj}
\lambda_j=\sqrt{\frac{j}{j+1}}\sqrt{\frac{n_q}{Q}} \mu_0(Q).
\end{equation}
The state~(\ref{piphi}) has neither a defined total angular momentum nor a good symmetry, but its is characterized by a parity $(-1)^K$. By combining states (\ref{piphi}) with the same value of $K$ (or $Q$), it is generally possible to build a physical state with good quantum numbers and good symmetry properties, but the task can be technically very complicated \cite{silv85}. 

The baryon mean square radius can be computed with the eigenstates (\ref{piphi}) corresponding to the mass spectrum $M_{u1}$, leading to~\cite{afm,wave}
\begin{equation}\label{rad}
\left\langle r^2\right\rangle=\left\langle\frac{1}{n_q}\sum_{i=1}^{n_q}(\vec x_i-\vec R)^2\right\rangle
=\left[\frac{Q}{n_q\mu_0(Q)}\right]^2.
\end{equation}
Since this result only depends on the quantum numbers via $Q$, and since a physical state must be a combination of eigenstates with the same value of $Q$, (\ref{rad}) is also valid for a physical state.

In order to check our calculations, an upper bound for the ground state of Hamiltonian~(\ref{geneh}) has been obtained by a variational method with a trial function $\phi$ given by (\ref{piphi}) with $K=0$ and new scale parameters $\lambda_j=\sqrt{j/(j+1)}\Lambda$, where $ \Lambda$ is a free quantity. By minimizing $\left\langle\phi\right| H\left|\phi\right\rangle$ with respect to $\Lambda$ (see \textit{e.g.} \cite{afm,gauss} for a computation of the needed matrix elements), this leads to the ground state upper bound
\begin{equation}
M^2_{u2}=\frac{16}{\pi}P(a_1+a_2\sqrt P)(2-b\sqrt P).
\end{equation}

Not only upper bounds can be obtained, but also a lower bound. Remarking that
\begin{equation}
\sum_{i=1}^{n_q}\left|\vec x_i-\vec R\right|>\frac{1}{n_q}\sum_{i<j=1}^{n_q}\left|\vec x_i-\vec x_j\right|
\end{equation}
thanks to the triangular inequality, eigenenergies of the Hamiltonian
\begin{equation}
H^I=\sum_{i=1}^{n_q}\sqrt{\vec p^{\, 2}_i+m^2}+\sum_{i<j=1}^{n_q}\left[\bar a\left|\vec x_i-\vec x_j\right|-\frac{b}{\left|\vec x_i-\vec x_j\right|}\right], 
\end{equation}
with $\bar a=a_2+a_1/n_q$, are lower bounds of the eigenenergies of Hamiltonian~(\ref{geneh}). Furthermore, a lower bound on the ground state mass of $H^I$ can be found in \cite{hall}, that is
\begin{equation}\label{ml1}
M_{l1}=n_q\, \inf_{\left|\psi\right\rangle}\left\langle \psi\left|\sqrt{\vec p^{\, 2}+m^2}+\frac{n_q-1}{2}\left[\bar a r-\frac{b}{r}\right]\right|\psi\right\rangle.
\end{equation}
More explicitly, an approximate formula for $M_{l1}$ at $m=0$ can be found using the AFM~\cite{afms} 
\begin{equation}\label{ml2}
 M_{l1}^2\approx M_{l2}^2=2 P (a_1+a_2 n_q) \left[2-b(n_q-1)\right].
\end{equation}
The variational character of $M_{l2}$ cannot be guaranteed, but since this last AFM approximation is very accurate~\cite{afms}, it can be reasonably supposed that $M_{l2}$ is still a lower bound of the ground state mass of~(\ref{geneh}). This has been numerically checked on several systems for $n_q=2$ and 3.

Since $\min (Q\, n_q) = 3 P$, one can verify that $M_{u1}(Q) > M_{u2} > M_{l2}$ for $n_q \ge 2$, as expected. The crucial point of these calculations is that the three bounds have the same behavior for large values of $n_q$. So we can have confidence on the large $n_q$-behavior of the exact solutions. 

\end{appendix}

\end{document}